# Fluid Simulation System Based on Graph Neural Network

Qiang Liu[a], Wei Zhu[b*], Xiyu Jia[c*], Feng Ma[d*], Yu Gao[e]

**Abstract**

Traditional computational fluid dynamics calculates the physical information of the flow field by solving partial differential equations, which takes a long time to calculate and consumes a lot of computational resources. We build a fluid simulation simulator based on the graph neural network architecture. The simulator has fast computing speed and low consumption of computing resources. We regard the computational domain as a structural graph, and the computational nodes in the structural graph determine neighbor nodes through adaptive sampling. Building deep learning architectures with attention graph neural networks. The fluid simulation simulator is trained according to the simulation results of the flow field around the cylinder with different Reynolds numbers. The trained fluid simulation simulator not only has a very high accuracy for the prediction of the flow field in the training set, but also can extrapolate the flow field outside the training set. Compared to traditional CFD solvers, the fluid simulation simulator achieves a speedup of 2-3 orders of magnitude. The fluid simulation simulator provides new ideas for the rapid optimization and design of fluid mechanics models and the real-time control of intelligent fluid mechanisms.

**Keywords**

Graph Neural Network    Adaptive Sampling    Cylinder Flow    CFD

* Beijing Institute of Technology, Beijing 100081, China

*E-mail address:* b)*wei.zhu@bit.edu.cn*

a)*3120200156@bit.edu.cn*

c)kang@bit.*edu.cn*

d)*3120150116@bit*.edu.cn

# 1. Introduction

Solving partial differential governing equations based on physical models is one of the important basic methods for human beings to explore, understand and simulate the real world. In the past 100 years, solutions (partial differential equations) have undergone a revolutionary leap from analytic approximation to numerical approximation. Especially since the 1970s, CFD technology represented by high-precision numerical algorithms has played a dominant role in solving complex and intense nonlinear flow problems such as shock waves, material discontinuities, and turbulence. CFD can help researchers profoundly and intuitively understand the details that are difficult to observe experimentally in complex fluid problems and quantitatively predict the development of complex fluid problems. It has become an indispensable and crucial auxiliary tool in human production activities. The differential governing equations are transformed into algebraic equations via discretization schemes, such as the finite difference or finite volume. In order to approach the solution of the original equation more accurately, the smaller the distance between discrete nodes is, the better the number of nodes used for interpolation is. Therefore, the number of nodes and the relationship between nodes are the key factors affecting the accuracy and efficiency of numerical calculation.

In the past 20 years, the leapfrog development of hardware-level has provided significant support for the popularization and large-scale application of CFD technology. However, the total reliance on hardware improvement to improve the overall level of CFD computing has reached a bottleneck. The core reason is that the innovation of algorithm theory has entered the stagflation stage: the last revolutionary numerical algorithm design idea -- TVD thought was put forward in 1978; The ENO scheme has broken the limit of finite-difference on the number of interpolation nodes. For example, the xx method has reached 132 order accuracy at most, but it has been 40 years since it was proposed, and the accuracy of finite volume and finite element algorithms is still challenging to break through 2~3 order. Particle algorithms (or meshless algorithms) have advantages when dealing with high-speed collisions of solids or crack generation and propagation, but they have never been mathematically proven.

From the perspective in the application, the role of traditional computational fluid dynamics in fluid engineering applications is to achieve quantitative characterization of flow through off-line analysis and law extraction to provide design, optimization, and control criteria for physics-based engineering of fluid systems. However, intelligent and agile fluid engineering applications represented by bionic fluid power require real-time acquisition and inversion of uncertain and incomplete complex flow laws in an open environment. Therefore, new technology is urgently needed to improve the real-time and dynamic inversion capability of flow field[1].

Generally speaking, computational fluid dynamics is based on the deduction of physical equations to obtain scientific laws, while machine learning can directly analyze, induce and deduce data to achieve law extraction. Machine learning uses probability, statistics, approximation theory, and complex algorithms to extract data relevance from massive complex data and achieve fast data reasoning and analysis capability under regular conditions[2]. It has strong complementarity with computational fluid mechanics and has the potential to surpass the computational method of fluid mechanics based on the physical model.

Since the beginning of the new millennium, with the explosive growth of machine learning technology represented by deep learning, many scholars are also trying to introduce deep learning methods into the fluid simulation. Some scholars train the neural network model through a large number of observation data and introduce the physical information in the data into the neural network by training a large number of data, namely, the observation bias into the neural network[3,4,5]. AliKashefi proposed a deep learning architecture for flow field prediction in irregular domains by using PointNet architecture[6]. Conducting neural networks by observing bias is the simplest method, but observational data is usually generated through expensive experiments or large-scale numerical simulations. In order to reduce excessive data dependence and improve the interpretability of the model, scholars introduced induction bias into neural network[7,8,9]. They embedded induction bias and prior knowledge related to a given task into the neural network to make the neural network have the properties related to a given task. Alvaro Sanchez-Gonzalez proposed a general framework based on data-driven simulation, "Graph-based Network Simulator", which introduced a strong inductive bias into the framework, where rich physical states are represented by interacting particle graphs, and complex dynamics are approximated by MLPS between nodes[10]. JuliaLing uses an invariant tensor basis to embed Galilean invariance into a network structure, which improves the prediction accuracy of neural networks in turbulence modeling[11]. Embedding inductive bias and prior knowledge into neural networks requires elaborate design, often challenging. Another method is to add learning bias to the neural network, giving the neural network prior knowledge by punishing its loss function. M.Raissi proposed the hidden fluid mechanics model[12], taking the simplified NS equation and transport equation as the loss function and making the neural network close to the NS equation and transport equation by punishing the loss function[13]. Observation bias, inductive bias, and learning bias can be used separately or in combination, and each bias can reinforce the other when used together.

Networks such as MLP and CNN have defects in processing irregularly structured data, so some scholars put forward the idea of graph theory [14] and graph convolution [15] to process irregularly structured data. Graph convolution introduces inductive bias into a neural network to process graph-structured data (irregularly structured data).In practical engineering, a fluid system is a multi-physical and multi-scale dynamic model, so the computational domain of computational fluid dynamics is usually irregular mesh, such as the flow field on the wing surface[16]. Inspired by this, we propose a graph machine learning model to deal with unstructured high-dimensional nonlinear flow field problems. The connection problem of Euclidean space is raised to a graph structure problem of non-Euclidean space using a graph machine learning model. Graph machine learning can define functions in non-Euclidean space and transfer high-dimensional information in the domain through the defined functions[17]. High-dimensional information can reduce the complexity of nonlinear relationships between data[18]. Therefore, we implemented our fluid simulation deep learning architecture through the graph attention network architecture[19] and found its applicability under different working conditions and different Reynolds numbers. Our fluid simulation simulator builds on previous work and presents the following innovations for different challenges:

- The key region of fluid computing is where the fluid physical gradient changes violently. In order to better capture gradient information and improve calculation accuracy, we use graph attention architecture. At the same time, there is a mutual coupling relationship among the physical quantities in the flow field, which indicates that the relations among physical quantities are diverse. We use the multi-headed attention mechanism to describe physical relationships from different presentation subspaces.

- How can we establish relationships between compute nodes and their neighbors to capture multi-scale physical information in fluid systems? Generally speaking, a fixed number of neighbor nodes will be selected to establish the neighbor node relationship, but large-scale physical information will be lost for multi-scale flow field data. Therefore, an adaptive sampling mechanism is proposed in this paper. By demarcating the appropriate scale range to find neighbor nodes, the graph neural network can capture multi-scale physical information to describe the details of the flow field more accurately.

- Flow field data's time and space dimensions are very large, and full-field, high-resolution and long-term simulation operation has high requirements and takes a long time for the computing platform, especially for the early training process of deep learning. In order to improve the training efficiency of the model and expand the scope of application of the model, we use the sampling-aggregation mechanism of graph neural network[20] to implement the super-resolution technology, train the fluid simulation model with low-resolution data, and apply the model to high-resolution fluid data at the same time.

## 2. The network architecture

**2.1 Figure the junction of neural network and computational fluid dynamics**

Navier-stokes equations are used in computational fluid dynamics to calculate the physical characteristics of the flow field with a strong coupling relationship. The physical quantities contained in Navier-Stokes equations are complex nonlinear coupling relations, leading to the difficulty of solving them. The universal approximation theorem[21] states that a sufficiently large and deep network can approximate any function. On the one hand, graph neural network has very high parameter complexity. On the other hand, graph neural networks and computational fluid dynamics have very high similarities in aggregation information. This provides a natural combination of graph neural networks and computational fluid dynamics. The relation between neural networks and computational fluid dynamics is shown below.

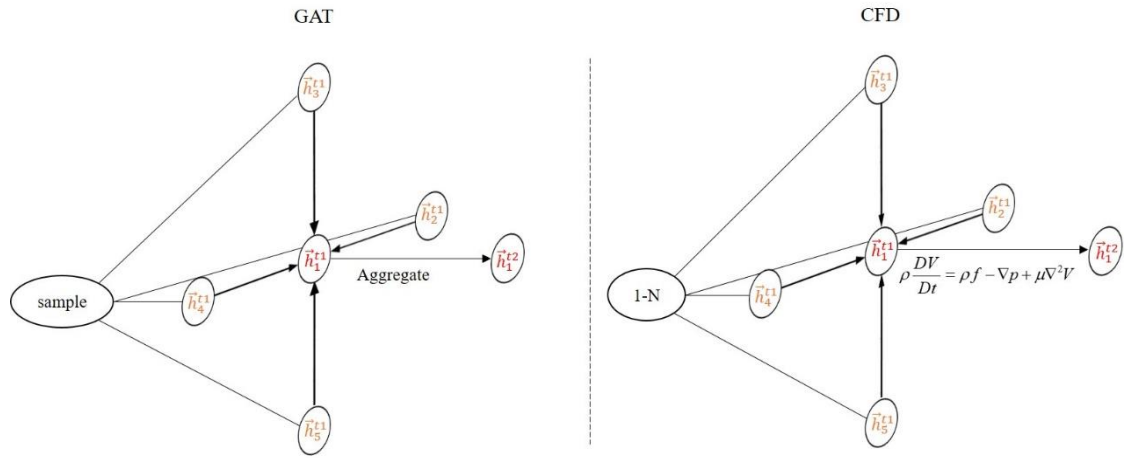

Figure 1: (left) Calculate flow field physical information by graph neural network. (right) Calculate flow field physical information by numerical calculation.

The right part of the figure is divided into graph neural networks, and the left part is computational fluid dynamics. Both operate directly on irregularly structured data. The graph neural network aggregates local information by sampling aggregation mechanism, and CFD solves Navier-Stokes equations by numerical method to transfer physical information between nodes.

Firstly, the graph neural network in this paper carries out message transmission through sampling and aggregation mechanisms. What sampling and aggregation methods do we choose specifically?

A fluid system is a multi-physical and multi-scale system. In computational fluid mechanics, people ensure the average of physical information cascade difference between two nodes to improve calculation accuracy. Therefore, the grids number is much larger near the wall or obstacles, where the gradient changes are more significant than the stable flow. Selecting first-order neighboring nodes or a fixed number of neighboring nodes will make the gradient difference between the compute node and its neighboring nodes too average so that the neural network cannot perceive the drastic gradient change in those regions. In this way, jitter will occur in the flow field with large gradient changes. Therefore, we consider sampling neighbor nodes adaptively according to the gradient variation law. Multi-scale gradient information is simultaneously aggregated by sampling many neighbor nodes in the region with large gradient changes so that the whole information is not lost while describing the details. The comparison between adaptive sampling and fixed neighbor node sampling is shown in the figure.

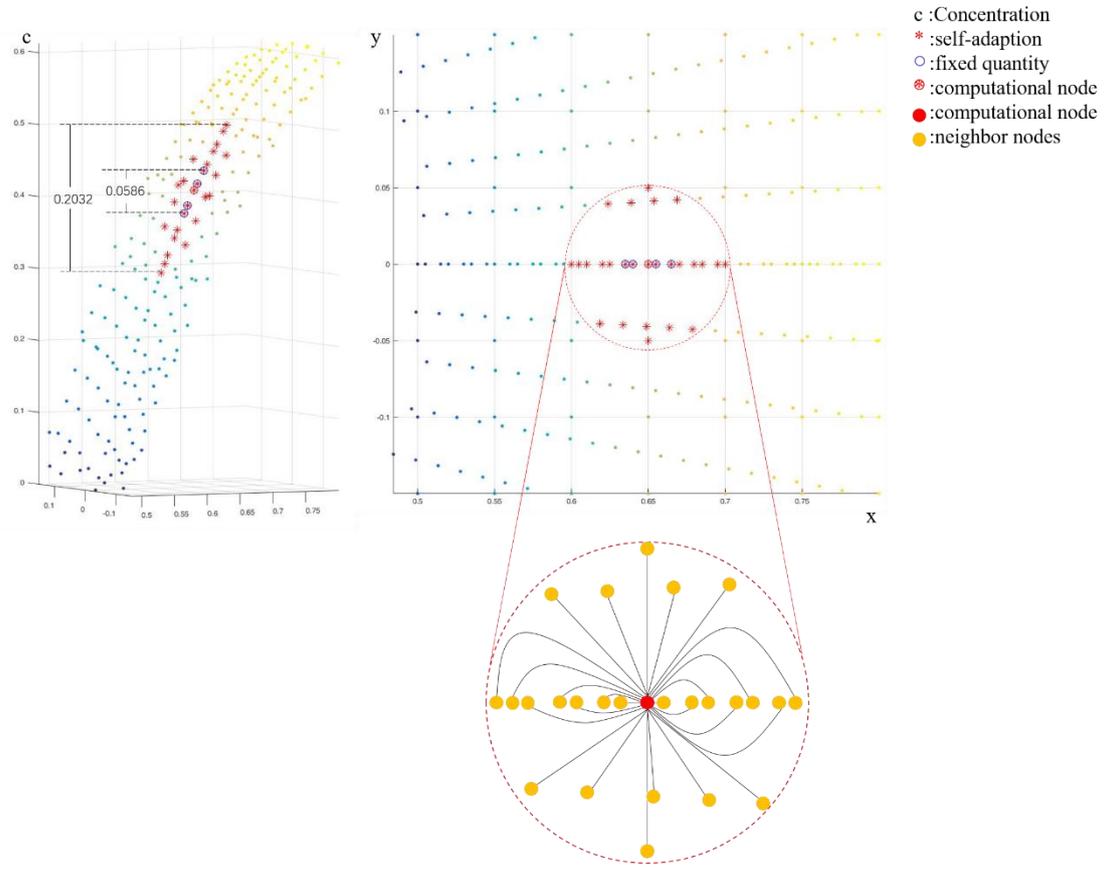

Figure 2: Adaptively select nodes within the sampling radius

The adaptive sampling neighbor node formula is as follows:

$$ne[i] = \{ j \mid j \in (x - x_i)^2 + (y - y_i)^2 \leq r \} \quad (1)$$

**2.2 Figure neural network architecture**

**2.2.1 Node symbol**

Taking the flow field around a cylinder as an example, we use the fully supervised learning method to train the model and learn the physical information of the flow field at continuous moments to achieve end-to-end physical information prediction. We establish the node neighbor relationship in the Spatio-temporal domain and implicitly embed the node coordinate information of the flow field into the input vector. The input state vector of each node is concentration, pressure, velocity in the X-axis direction, and velocity in the Y-axis direction at six continuous moments. The input vector of a single node is:

$$x_i^{t_1-t_6} = [c_i^{t_1-t_6}, p_i^{t_1-t_6}, u_i^{t_1-t_6}, v_i^{t_1-t_6}]$$

The overall input vector represents:

$$x^{t_1 \sim t_6} = [c^{t_1 \sim t_6}, p^{t_1 \sim t_6}, u^{t_1 \sim t_6}, v^{t_1 \sim t_6}, e]$$

Where e represents node connection, c represents concentration, p represents pressure, u represents velocity along the X-axis, and v represents velocity along the Y-axis.

The output vector of the node is:

$$y_i^{t_7} = [c_i^{t_7}, p_i^{t_7}, u_i^{t_7}, v_i^{t_7}]$$

The overall output vector is:

$$y^{t_7} = [c^{t_7}, p^{t_7}, u^{t_7}, v^{t_7}, e]$$

In the simulator, the input vector $x^{t_1 \sim t_6}$ is used to deduce the output vector $y_i^{t_7}$, and the simulator function is expressed as follows:

$$y_i^{t_7} = f(x_i^{t_1-t_6}, \sum_{j \in ne[i]} x_j^{t_1-t_6}, e_{ij}, \omega) \qquad (2)$$

There $ne[i]$ represents the neighbor node of node i, $e_{ij}$ represents the connection edge of i and j, and $\omega$ represents the learnable parameter.

**2.2.2 The network architecture**

The parameterized function $f(\omega)$ computes fluid physical information for $x \to y$. We use the graph attention layer to construct the network architecture, and the overall network architecture is shown below.

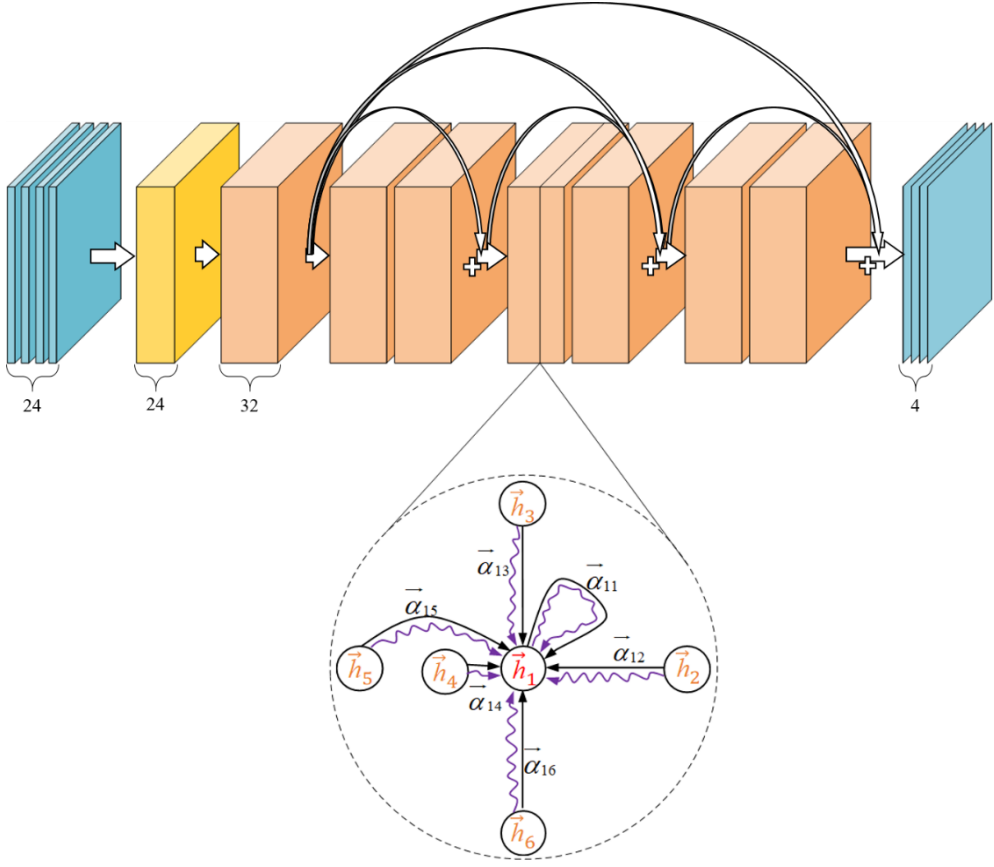

Figure 3: The network architecture built by the graph attention layer. The normalization layer dimension is 24. The dimension of the graph attention layer is 32. For the specific network architecture, see Section 3.

The network architecture is composed of a single-headed attention layer and a multi-headed attention layer. The attention layer uses the attention mechanism to establish the weight matrix of the relationship between computing nodes and neighboring nodes and aggregate the physical information of neighbors, as shown in the figure above. In fluid calculation, pressure, velocity, density, etc., are strongly coupled. Our model's input and physical output quantities include concentration, pressure, and velocity. In Navier-Stokes equations, pressure and velocity are nonlinear, and the parameter matrix of attention coefficient characterizes the nonlinear relationship between node features. At the same time, the weight between nodes is allocated by the attention coefficient, which is calculated as follows:

$$q = \sum_{j \in ne[i]} \vec{h}_j \times W_q$$

$$k_i = \vec{h}_i \times W_k$$

$$v_i = \vec{h}_i \times W_v$$

$$\alpha_{ij} = \frac{\exp(s(\sigma(k_i), \sigma(q_j))/\sqrt{d})}{\sum_{k \in ne[i]} \exp(s(\sigma(k_i), \sigma(q_k))/\sqrt{d})} \quad (3)$$

there $W_q$, $W_k$, $W_v$ represent the linear transformation matrix corresponding to the query vector, keyword vector, and value vector, $\vec{h}_j$ represents the neighbor node, $\vec{h}_i$ represents the compute node, $d$ represents the dimension of the query vector and keyword vector, $s$ represents the dot product, $\sigma$ represents the LeakyReLU activation function, and $\alpha_{ij}$ represents the attention coefficient.

Because the model has multiple inputs and multiple outputs, there is a complex coupling relationship between inputs and outputs. In order to describe the correlation between nodes more comprehensively within the network architecture. In the network architecture, we combine a single-headed attention mechanism with multi-headed attention mechanism. Single-headed attention mechanism couplings node feature relations, and multi-headed attention mechanism can describe node relations more comprehensively. The multi-attention mechanism is shown below.

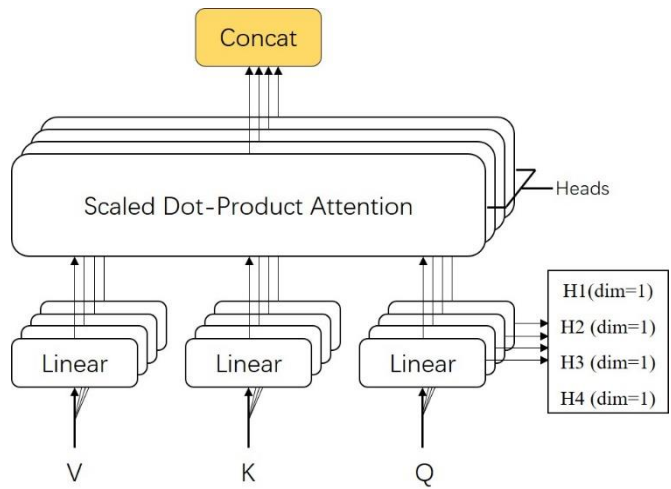

Figure 4: Multi-head attention mechanism

The multi-headed attention mechanism puts V, K and Q vectors into four representation subspaces, respectively, and the representation subspace dimension of K and Q is 1.

In order to improve the complexity of the network model, on the one hand, it is necessary to increase the width of the network (i.e. the dimension of each layer of the network), and on the other hand, it is necessary to overlay the network depth. Meanwhile, a k-layer graph convolutional neural network can fuse the information of k-order neighbor nodes. Therefore, the deep network has a wider receptive field and can capture farther physical information. The following figure shows the aggregation node information of the two-layer graph neural network.

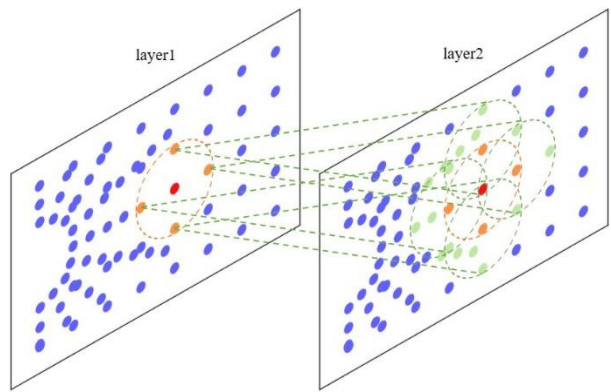

Figure 5: The receptive field of the network continues to expand through the superposition of multiple layers.

In the process of transferring physical information to the deep network, the information of the shallow network will be gradually lost. In order to transfer the shallow network information to the deep network and ensure the complexity of the network structure, we add residual connections into the network architecture. The method is to add a residual connection between the multi-headed attention and the single-headed attention layers and add residual connection in the first layer of the neural network and every two subsequent layers.

The formula of the single-headed attention layer and the formula of the multi-headed attention layer are shown as follows:

$$\vec{h'}_i^l = \sigma(\sum_{j \in ne[i]} \alpha_{ij} \cdot (\vec{h}_j^l \times W^l)) \qquad (4)$$

$$\vec{h'}_i^l = \sigma(\frac{1}{K}\sum_{k=1}^{K} \sum_{j \in ne[i]} \alpha_{ij}^k (\vec{h}_j^l \times W^l)) \qquad (5)$$

The residual connection formula between layers is as follows:

$$y_l = h(x_l) + F(x_l, W_l) \qquad (6)$$

$$x_{l+1} = \sigma(y_l) \qquad (7)$$

Where $x_l$ and $y_l$ represent the input and output of layer $l$ respectively, $F$ represents the attention layer function, $h(x_l)$ is the identity mapping of $x_l$, and $\sigma$ is the LeakyReLU activation function. Based on the above formula, we can obtain the learning characteristics from shallow $l$ to deep $L$ as follows:

$$x_L = \sigma(x_l + \sum_{i=l}^{L-1} F(x_i, W_i)) \qquad (8)$$

# 3. Results

**The result details**

We added BN layer to eliminate gradient explosion before the attention layer and set the dropout=0.6 and Learning rate=0.001 between layers. Use the mean square deviation as the loss function. For 2D flow field calculation, the dimension of the middle layer is set as 32, and the dimension of the output layer is set as 4. For 3D flow field calculation, the dimension of the middle layer is set as 64 and the dimension of the output layer is set as 5.

There are altogether 8 graph attention layers in the network architecture, and the number of heads of each layer, dimension and activation function of Q and K vectors are shown in the following table

| Layer | Heads | Dimensionality | Activation Function |
|---|---|---|---|
| 1 | 4 | 4 | Leaky_ReLU |
| 2 | 1 | 16 | Leaky_ReLU |
| 3 | 4 | 4 | - |
| 4 | 1 | 16 | Leaky_ReLU |
| 5 | 4 | 4 | - |
| 6 | 1 | 16 | Leaky_ReLU |
| 7 | 4 | 4 | - |
| 8 | 1 | 1 | - |

Table 1: Network Architecture Details

We implemented our model using TensorFlow, tf_Geometric[22], and SKLearn. Hardware devices use NVIDIA TITAN RTX2070 and Intel(R) Xeon(R) CPU E5-2660 V4 @ 2.00GHz.

We compare the accuracy of different network architectures on 2D circular cylinder flow data with Re=100 and Pec=100.In addition, 3D circular cylinder flow data around a cylinder with Re=100 and Pe=100, 2D circular cylinder flow data around a cylinder with Re=200 and Pec=2000 under Newman boundary condition, 2D circular cylinder flow data around a cylinder with Re=200 and Pec=2000 under

Dirichlet boundary condition, Re=1.0/0.0101822,The 3D aneurysm data with Pec=1.0/0.0101822 and the flow data with Re=5 and Pec=15 tested the generalization ability of our model. These data come from Brown University's Department of Applied Mathematics[13].

### 3.1 Fluid simulation simulator

The design ideas and network architecture of the fluid simulation simulator are explained in section 2, which verifies the simulation ability of the fluid simulation simulator and the influence of different network designs on the prediction results through experiments. Fluid simulation model prediction fluid data flow is shown in the figure below.

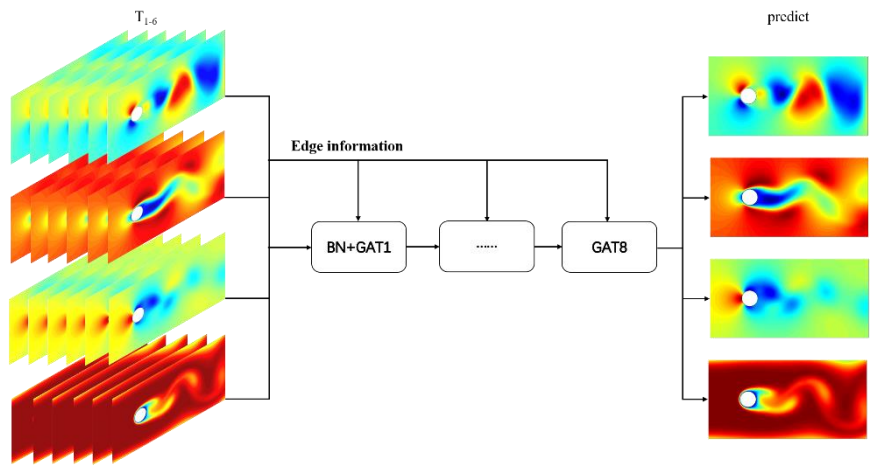

Figure 6: The simulator extrapolates the physical quantities of the seventh moment through the physical quantities of six consecutive moments.

### 3.2 Calculation precision analysis of different network architectures

We show the computational accuracy of the network through the relative error. The figure above shows the prediction accuracy of the flow field around the cylinder of different models in 16 seconds.The comparison of the table shows that our network architecture has significant advantages in computing accuracy compared with other network architectures. GraphSAGE is the average aggregation without attention mechanism and has the worst effect. In terms of sampling, the effect of sampling with fixed number of neighbor nodes is inferior to that of self-using neighbor nodes. In terms of network architecture, the effect of the network architecture with no residual connection and single-headed attention mechanism is worse than our network architecture, indicating that the residual connection and multi-headed attention mechanism have excellent effects. Compared with HFM, our network has a significant advantage in velocity and pressure prediction accuracy, while concentration prediction accuracy is slightly inferior to HFM. At the same time, GCN using residual connection and adaptive sampling also achieved good results.

| Methods | L2 relative error | | | |
|---|---|---|---|---|
| | C | U | V | P |
| Ours | 0.0223 | 0.0129 | 0.0128 | 0.0148 |
| No Residual | 0.0291 | 0.0194 | 0.0251 | 0.0215 |
| Self-Attention | 1.1823 | 0.7319 | 0.3870 | 0.1351 |
| Fix-Point | 0.0376 | 0.0215 | 0.0208 | 0.0184 |
| GCN | 0.0507 | 0.0404 | 0.0349 | 0.0430 |
| GraphSAGE | 4.7874 | 0.0606 | 1.0016 | 1.0000 |
| HFM | 0.0153 | 0.0680 | 0.0701 | 0.0987 |

Table 2: Comparison of the results of different methods

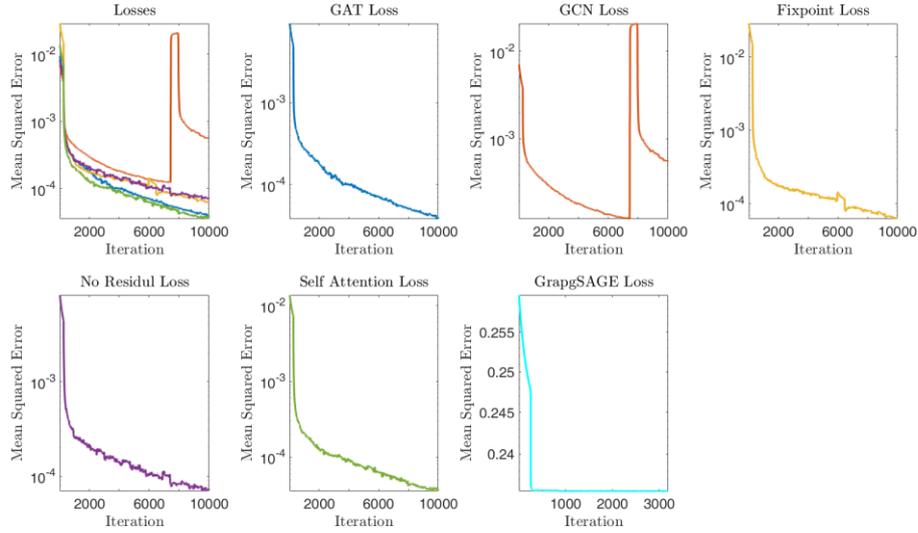

Figure 7: According to the Loss curve, the loss function of our network architecture is slightly inferior to the single-head attention model, but superior to other models. The loss function of the single-headed attention model is small, but the relative error is large, indicating that the single-headed attention model has the phenomenon of over-fitting.

At the same time, we compared the accuracy of network architecture with different layers, and found that concentration prediction accuracy was optimal at 10 layers, speed prediction accuracy at 4 layers along the X axis, and speed and pressure prediction accuracy at 8 layers along the Y axis. Overall consideration, we choose 8-layer network architecture as the experimental model.

| Layers | L2 relative error | | | |
| --- | --- | --- | --- | --- |
| | C | U | V | P |
| Ours(8-layer) | 0.0223 | 0.0129 | 0.0128 | 0.0148 |
| 2-layer | 0.0275 | 0.0151 | 0.0159 | 0.0168 |
| 4-layer | 0.0221 | 0.0125 | 0.0149 | 0.0164 |
| 6-layer | 0.0246 | 0.0168 | 0.0159 | 0.0152 |
| 10-layer | 0.0213 | 0.0146 | 0.0136 | 0.0140 |
| 12-layer | 0.0229 | 0.0127 | 0.0155 | 0.0159 |

Tabel 3: Comparison of results with different network layers

**3.3 The spatial distribution of errors**

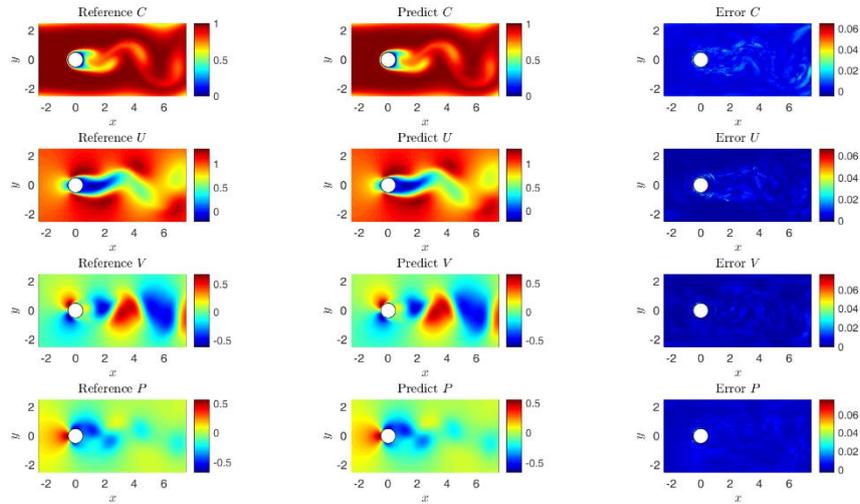

Figure 8:First of all, the figure above shows the comparison between the predicted data of the flow field around the cylinder at the 100th moment and the simulated data of the true value when Re is 100 and Pe is 200.By comparing the physical quantities C, U, V and P in the figure, it is proved that the model can capture the physical laws of the flow around a cylinder.

Next, we show the absolute error distribution of physical quantities in the flow field in the form of a two-dimensional thermal diagram. It can be seen from the figure above that the absolute error of four physical quantities C, U, V, and P in the flow occurrence zone and flow zone is large. The area with the largest error is the junction of the high-density grid and low-density grid. Nodes in the high-density grid region can express more precise physical information, while the physical quantity with drastic gradient changes loses certain details from the high-density grid to the low-density grid, resulting in the loss of calculation accuracy. Combined with the predicted data, it can be seen that the gradient of concentration and velocity in the X-axis direction changes the most at the junction, with more details. However, the change of pressure and velocity gradient in the Y-axis direction is small, and it is more large-scale physical information.

**3.4 Time variation rule of error**

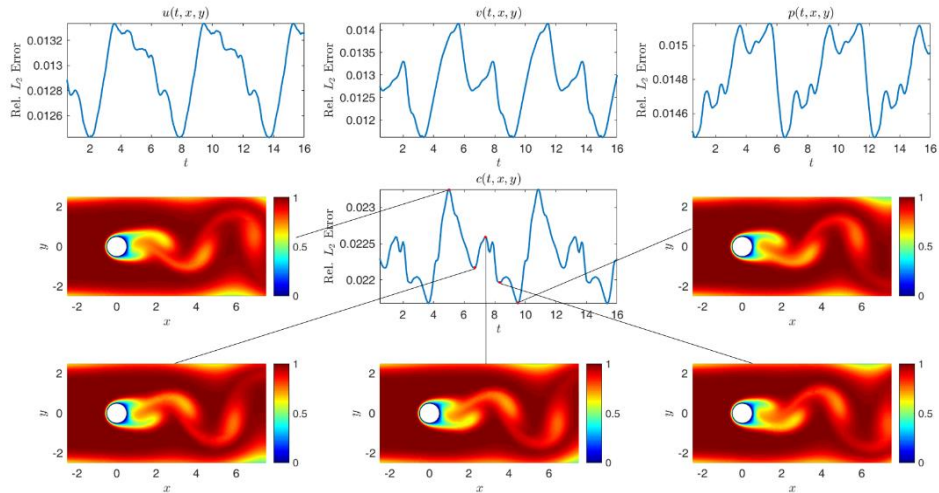

Figure 9:The figure above shows the curve of the average L2 relative error of the flow field at 201 moments (16 seconds). The concentration error curve shows the flow field information corresponding to the error peak value and valley value within a period. It can be seen from the flow field information graph that the relative error is the largest when the physical information change area is the largest in the flow field; conversely, the relative error is the smallest when the physical quantity gradient change area is the least. Therefore, the gradient change in the flow field impacts the calculation accuracy.

**3.4 Network generalization ability test**

Next, we tested the generalization ability of the network architecture. First, we tested the flow field around a cylinder under different Pec and boundary conditions.

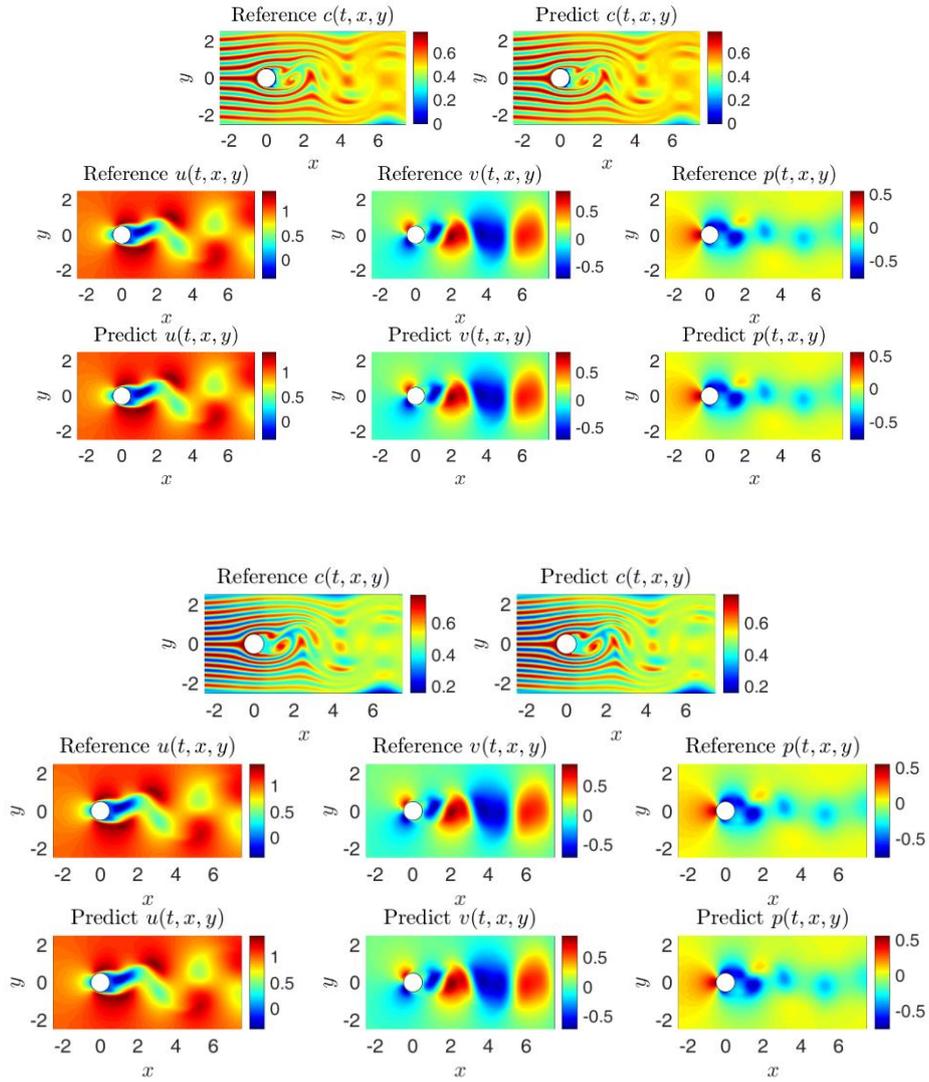

Figure 10: The figure above shows the comparison between the predicted results of Dirichlet boundary condition and Newman boundary condition and truth-value simulation data under Re=200 and Pec=2000, respectively. The following table shows the prediction accuracy of these two working conditions.

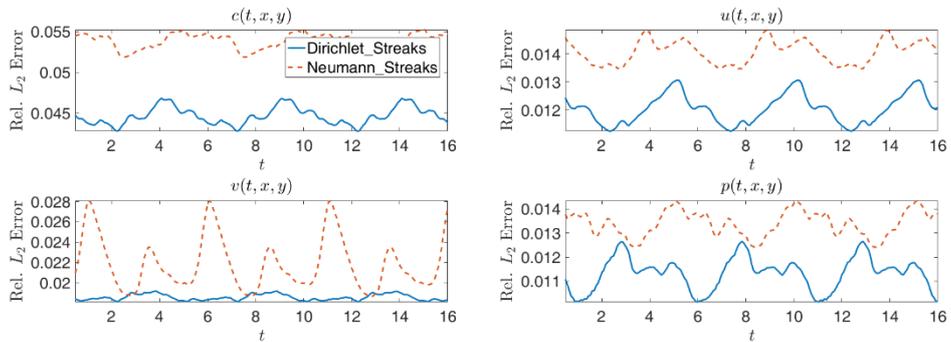

|  | L2 relative error | | | |
| --- | --- | --- | --- | --- |
| Working condition | C | U | V | P |
| Dirichlet Streaks | 0.0538 | 0.0141 | 0.0220 | 0.0134 |
| Neumann Streaks | 0.0447 | 0.0120 | 0.0186 | 0.0113 |

Figure 11: For these two kinds, the prediction accuracy of concentration is the worst, and the prediction accuracy of pressure is the best, which shows that the prediction accuracy can be improved through mutual coupling between velocity and pressure from another aspect.

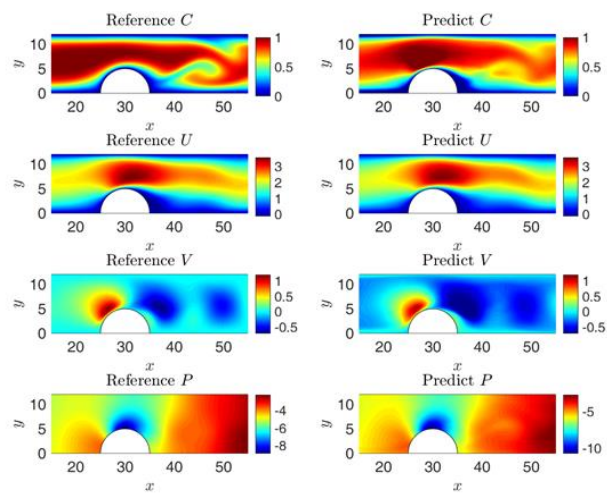

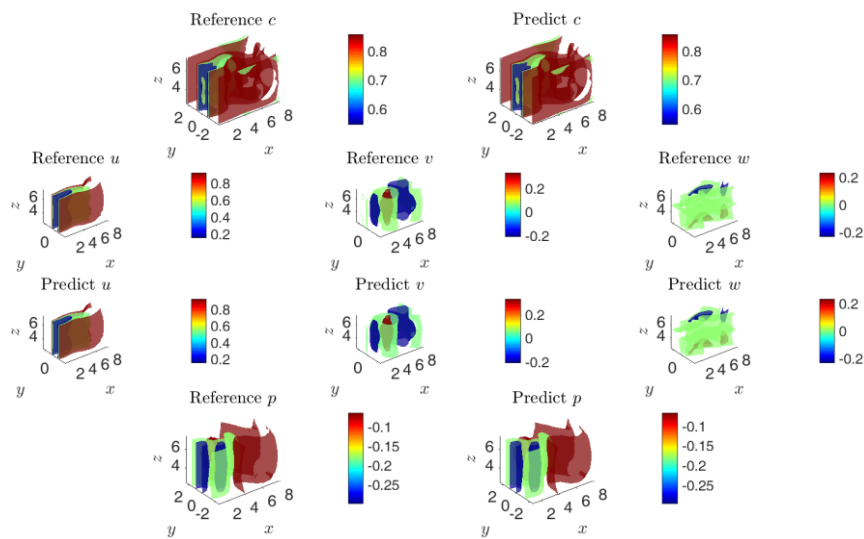

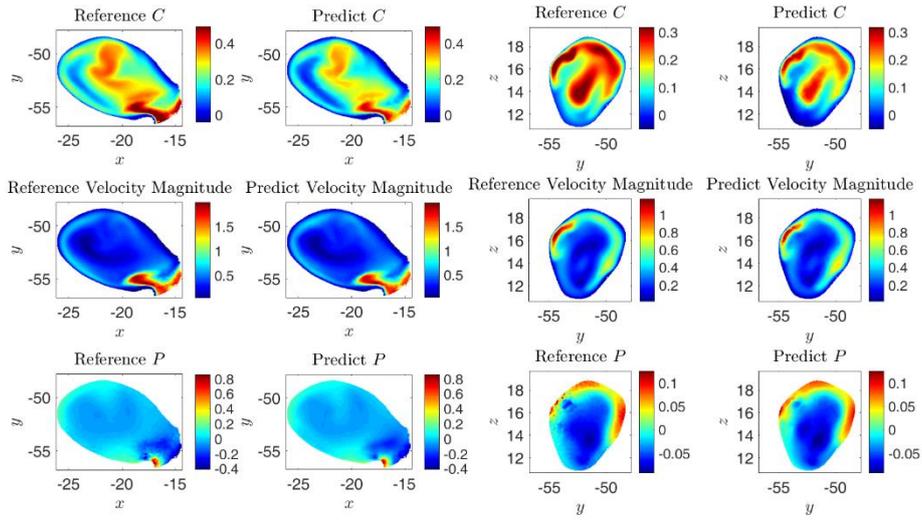

Figure 12: The figure above compares predicted results and true value simulation data of semi-cylindrical flow, three-dimensional cylindrical flow and three-dimensional aneurysm flow. It can be seen that the prediction results of semi-cylinder flow around and THREE-DIMENSIONAL aneurysm flow can well grasp the overall information, but the simulation ability of subtle information is poor.

| Working condition | L2 relative error | | | | |
|---|---|---|---|---|---|
| | C | U | V | W | P |
| Stenosis2D | 0.3082 | 0.1709 | 0.3387 | - | 0.0432 |
| Aneurysm3D | 0.3222 | 0.1755 | 0.1481 | 0.2034 | 0.2820 |
| Cylinder3D | 0.0277 | 0.0103 | 0.0120 | 0.0284 | 0.0225 |

Table 4: L2 relative error of different experiments

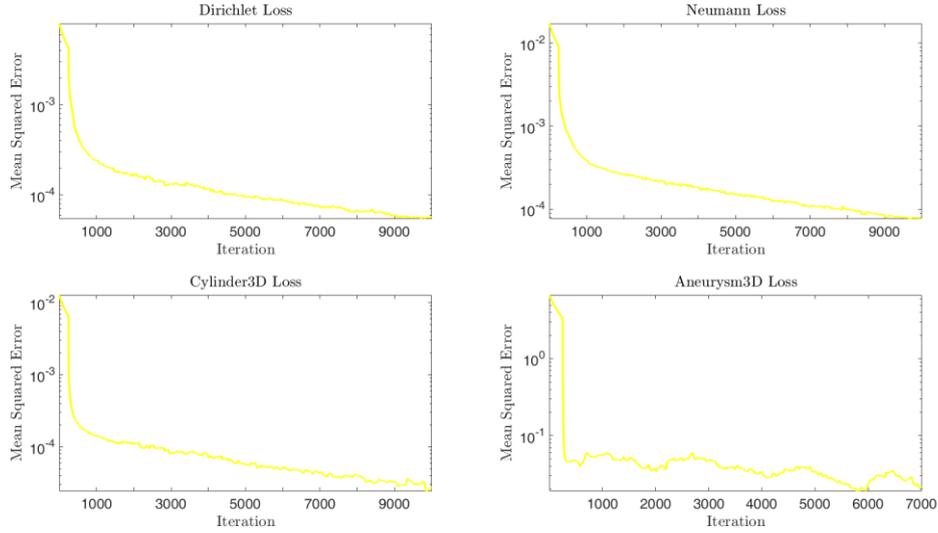

Figure 13: Loss functions for different experiments

**3.4 Super-resolution**

The "sample-aggregation" mechanism introduced in Section 2.2 gives the model a natural advantage in computing structural diagrams of different sizes. The "neighbor-aggregation" mode of message transfer is based on graph topology. The model learns how the neighbors of nodes transfer messages. The model parameters are independent of graph size and the number of nodes in the structure graph. Through spectral analysis of the model trained by original data, we found that a small amount of low-frequency information in the flow field data can grasp the overall physical law of the flow field, while high-frequency information is more about describing the details of the flow field. Therefore, we consider adding super-resolution technology to the physical model. In the dense grid area near the cylinder, we eliminate the nodes that account for 50% of the grid nodes in the dense area. The number of nodes after down-sampling accounts for about 78% of the original number of nodes. After down-sampling, the flow field data is used to train the graph neural network architecture. The experimental model is still used in the original flow field, and results are obtained.

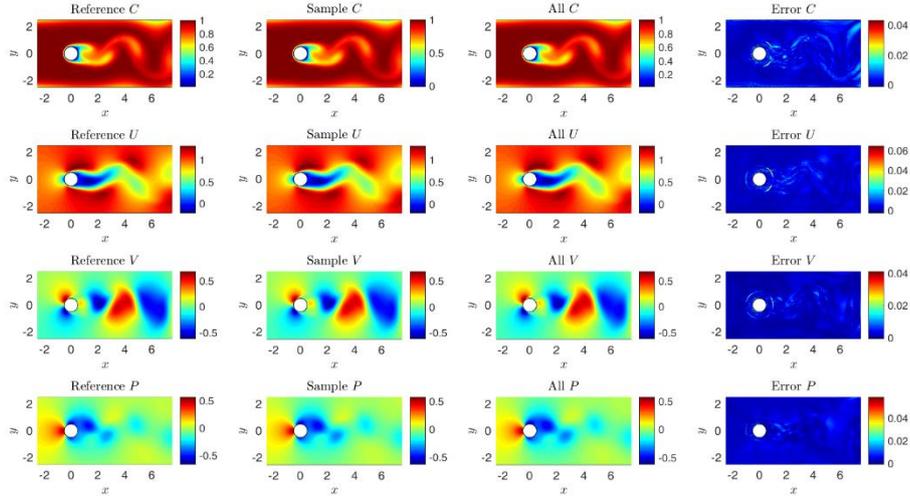

| Methods | L2 relative error | | | |
|---|---|---|---|---|
| | C | U | V | P |
| All | 0.0196 | 0.0154 | 0.0140 | 0.0157 |

Figure 14: The figure above shows the L2 relative errors tested on the original data of the model obtained by down-sampling data training. It can be seen that the high-resolution flow field data predicted by the down-sampling model still has high accuracy.

(1)

## 4. Conclusion

We propose a fluid simulation simulator with inductive bias and construct the fluid simulation simulator through a graph attention neural network. For multi-physical and multi-scale fluid systems, we use adaptive sampling and attention mechanisms to aggregate physical information of compute nodes and neighboring nodes. It is found that the adaptive sampling method is superior to the sampling method with a fixed number of neighbor nodes by comparing the experimental structure of adaptive sampling and the sampling method with a fixed number of neighbor nodes. In order to explore the importance of attention mechanisms, we compare the accuracy of average polymerization and attention mechanism polymerization in cylindrical flow experiments. Experimental results show that the ability of the attention mechanism to aggregate physical information of adjacent nodes is much higher than that of average aggregation. For the complex physical relations between nodes in the fluid system, we use a multi-head attention mechanism to learn node relations in different representation subspaces and describe the physical relations between adjacent nodes in a more diversified way.

As the graph neural network expands with the superposition of the network depth, the network can capture physical information at a larger scale to improve the network sensitivity field. We overlay the multi-layer graph attention layer to increase receptive field, and the 8-layer network architecture has the highest complete accuracy verified by experiments. At the same time, some shallow information will be lost when fluid information is transferred to a deep network in a graph neural network. We transfer the shallow information to the deep network through residual connection and obtain good results through experimental verification. Our network architecture is superior to the GCN layer in terms of concentration, pressure, and speed prediction. Compared with HFM, it is superior in pressure and speed and inferior in concentration.

We verify the generalization ability of the fluid simulation simulator in Dirichlet boundary and Newman boundary conditions of two-dimensional flow around a cylinder, two-dimensional half-cylinder flow field, three-dimensional cylinder flow field, and three-dimensional aneurysm flow field. We analyze the prediction accuracy in time series and find that the gradient in the flow field is negatively correlated with the prediction accuracy. The greater the gradient in the flow field, the lower the prediction accuracy, and the computational accuracy of computational fluid dynamics is consistent.

We now introduce inductive bias into the fluid simulation simulator, and in the future, we expect to introduce learning bias into the fluid simulation simulator. Further, improve the simulator's prediction accuracy, reduce the network's training time, and make the simulator more explanatory.